\documentclass[12pt,a4paper]{article}

\newcommand{\dd}{\mathrm{d}}
\newcommand{\dbar}{\;\,\bar{}\hspace{-5pt}\dd}

\newcommand{\etc}{\textit{et al}}
\newcommand{\comments}[1]{}

\begin{document}

\title{How and in What Sense Can the Entropy Be Measured?}
\date{August 9, 2002}
\author{Bin Zhou\thanks{E-mail address: zhoub@ihep.ac.cn} \\
   Department of Physics, Beijing Normal University \\
   Beijing 100875, P. R. China \\
   and \\
   Institute of High Energy Physics, Chinese Academy of Sciences \\
   P. O. Box 918-4, Beijing 100039, P. R. China}

\maketitle

\begin{abstract}

In this paper, a method of measuring the entropy is presented. Problems related
to the entropy and the heat are also discussed.

\end{abstract}


\section{Introduction}
\label{sec:Introd}

Among the branches of classical physics, thermodynamics and statistical physics
should be the most fantastic ones, which have attracted so many great minds
to devote their lives and energy to investigating every topic in these fields.
Perhaps among the branches of classical physics, they are the only ones that
leave so many open questions, not only in the field of application, but also
in their foundations.

In the last century, most of the physicists were attracted to the quantized
physics, which can be viewed as the opposite side of classical physics.
Although quantization has been the principal melody of physics ever since then,
thermodynamics and statistical physics are still vivid with their own open
questions. And, even in the so-called modern physics, shadows of thermodynamics
and statistical physics are also seen: In the black hole theory, black holes
are endowed with the temperature and the entropy\cite{blackhole}. Ideas and
concepts of thermodynamics and statistical physics are also applied to the
condensed matter theory and the string theory\cite{string}.
In fact, the history of quantum theory can be traced back to thermodynamics
and statistical physics, where M. Planck use the famous assumption $E=h\nu$
to derive Planck's formula\cite{Planck}, the spectral energy distribution of
black bodies.

On the other hand, there are so many open questions in thermodynamics as well
as statistical physics --- there are even no universal theories, neither
thermodynamic nor statistical, for non-equilibrium systems. Thus the
foundations of thermodynamics, which are claimed to be
valid for arbitrary thermal systems and arbitrary processes, could not be
tested by comparing the experimental data with the predictions from a universal
thermodynamics. This sounds not so nice to the society of physics. 

It will benefit us greatly if we exam the foundations of thermodynamics
carefully, no matter theoretically or experimentally. For example, if we were
able to measure the entropy of a system in an arbitrary given state, it would
be a piece of good news. However, whenever we try to fulfill such a scheme, we
always find that there are some confusing points or experimental obstacles in
thermodynamics, standing in the way to prevent us from approaching to our goal.

In this paper, a method is given as how to measure the entropy of a system in
an arbitrary given state. In order to simplify our discussion, we assume it to
be a gas system consisting of only one kind of molecules, with equilibrium
states being labeled by its temperature $T$, its volume $V$ and the amount
$n$ of its molecules measured in the unit of moles. As implied, such a system
is an open system.

According to thermodynamics, a reversible process is adiabatic if and only if
the entropy of the system is ``conserved", or, invariant, in the process. This
statement is theoretically explicit, but it is hard to be applied in
experiments. Usually a process is judged to be adiabatic empirically
or \textit{a priori}. In a thermodynamic measurement, one often wants to assure
that a process is adiabatic, or that certain a system, which may consist of
the system to be measured as well as some part of its surroundings, is
thermally isolated from other surroundings. Hence, in thermodynamics,
the error in a measurement concerning heat or thermodynamic functions are
hard to be estimated quantitatively.

The situation becomes rather worse in the case of open systems because the
concept of an adiabatic process is much far away from our  heuristic ideas.
Whenever the amount of molecules is changed in a reversible process, the
entropy accompanying the molecules that is brought into or out of the open
system accounts for certain part of the change of the entropy of the open
system. As a matter of fact, an adiabatic reversible process, namely, a
reversible process which remains the entropy invariant, will be peculiar from
the point of view of empiricism: A reversible process that seems not to
absorb or emit heat is not adiabatic provided the amount of molecules is
changing. The detailed analysis is presented in \S \ref{sec:Heat},
implying that there are serious problems in the theory of thermodynamics.
But this will not be the topic of this paper.

The concept of heat thus is made confusing. It arises to be a question whether
we can measure the heat correctly if one of the amounts of the components of
a system is changed in a process. If there is no problem, then the method in
this paper can be applied to measure the entropy of a system state by state.

This paper is organized as in the following. In \S \ref{sec:Problems}, some
points that might cause confusion or arguments are made clear first. Especially,
the entropy of an open system, as well as the heat absorbed by it, is analyzed.
It seems that the theory of thermodynamics of reversible processes is not
compatible with the thermodynamic laws. It is indeed the case\cite{Zhou}, but
this is not the topic of this paper.
In \S\ref{sec:Method}, we first outline of the method of how to measure the
entropy of an equilibrium gas system. Then the formulas are derived in details.
In \S\ref{sec:cd}, conclusions and further discussions are given.

For those who are interested in the measuring methods, \S\ref{sec:Outline} is
what the wanted. Those who want to see the detailed derivation may find it in
\S\ref{sec:ProposalExp}. If there is any question about these contents, then
\S\ref{sec:Problems} is referred to.

\section{Problems That Should Be Made Clear}
\label{sec:Problems}

Before we start to discuss how the entropy could be measured, some problems
should be made clear because the author astonishingly finds that there are so
much misunderstanding when the topic of this paper is discussed. Those who does
not want to quarrel with me, when he or she is reading the next section, may
safely skip the following contents in this section. But when he or she feels
eager to quarrel with me, I beg him or her to read this section carefully
before he or she comes up to me.

In this paper, we only discuss the open systems consisting of one kind of
molecules, with the amount denoted by $n$ in the unit of moles. For simplicity,
we assume that the system is a fluid system.

First let me emphasize that the correctness of thermodynamics of reversible
processes should never be questioned in this paper, because I am trying to
give some possible experimental methods, in its framework, of how to measure
the entropy of a system in an equilibrium state. From the discussions, it can
be revealed that there are some serious questions rooted deeply in
thermodynamics.  However, in order for the outline of this paper to be clear,
the correctness of thermodynamics should never be suspected.

\subsection{The Absolute Entropy and the Third Law of Thermodynamics}
\label{sec:AbsEntropy}

Perhaps someone wants to remind me that, in thermodynamics, functions such as
the internal energy $U$ and the entropy $S$ can be determined only up to a
constant, hence it should be suspected immediately that the absolute entropy,
while not the change of it, could be measured.

However, one should recall that it is not the case, at all.
When any one of the thermodynamic potentials such as the internal energy $U$,
the enthalpy $H = U + pV$, the Helmholtz free energy $F = U - TS$ or the Gibbs
free energy $G = U + pV - TS$, \etc, is assumed to be additive and to be a
homogeneous function of the first degree in its corresponding extensive
variables, the arbitrariness of all the constants in these potentials and the
entropy function ends up. For example, when $G(T,p,n)$ is assumed to be a
homogeneous function of the first degree in $n$, it can be obtained
that\cite{Sommerfeld,Zemansky}
\begin{displaymath}
  G = \mu n
\end{displaymath}
by using Euler's theorem for homogeneous functions. Hence $U$, $H$ and $F$ are
all obtained as in the following:
\begin{eqnarray}
  & & U(V,S,n) = - pV + TS + \mu n,
\label{Uidentity} \\
  & & H(p,S,n) = TS + \mu n,
\nonumber \\
  & & F(T,V,n) = - pV + \mu n.
\nonumber
\end{eqnarray}
Since $\dd F = - p\,\dd V - S\,\dd T + \mu\,\dd n$, the entropy
\begin{equation}
  S = -\bigg(\frac{\partial F}{\partial T}\bigg)_{V,n} = S(T,V,n)
\label{Sdef}
\end{equation}
is also determined\footnote{
To be more confirmative, suppose that $S' = S + S_0$ is another possible entropy
for the same system in the same state, with $S_0$ a constant. Then we obtain
another Gibbs free energy $G' = G - TS_0 = \mu n - T S_0$, which is no longer
a homogeneous function in $n$ whenever $S_0 \neq 0$. Constants in other
functions can be similarly discussed.
}, being a homogeneous function of the first degree in $V$ and $n$. Such an
entropy is what is suggested to be measured in this paper.

Someone may argue that this would make the third law of thermodynamics
unnecessary. Is that true? Of course, no, because this is not the business that
third law of thermodynamics can do. There are other conclusions, such as that
the absolute zero of temperature cannot be reached by any finite process, that
are not available without the third law.

Since there are various versions of the third law in literature, it is not
strange that there are lots of gentlemen who think that the third law can be
used to determine the integral constants in entropy\footnote{
  For example, it is written in the famous book\cite{LDEM} of L.~D.~Landau and
E.~M.~Lifshitz, saying that the entropy of any system vanishes at the absolute
zero of temperature. According to A. Sommerfeld, however, W. Nernst disliked
the concept of entropy\cite{Sommerfeld}. If so, it seems that the description
of Nernst's theorem in \cite{LDEM} is not a faithful version of Nernst's
original statement. And, whatsoever speaking, Sommerfeld's description is the
representative of the version that is widely accepted, although, in spirit, it
is equivalent to Landau and Lifshitz's statement in \cite{LDEM}. As for the
reason why they are equivalent, see the next two paragraphs.
}. Here we state the third law as that in \cite{Sommerfeld}: For a system, the
limit $\displaystyle{ S_0 = \lim_{T\to 0}S }$ is a constant value that is
independent of any variables such as the volume $V$, the pressure $p$ and the
amount $n$, \etc. It is in this sense that the determination of $S_0$ is not
the business of the third law only.

The third law of the thermodynamics concerns only the asymptotic behavior of
the entropy. It is a relatively independent thing to determine the constant
$S_0$ for a system. Unfortunately, in the literature it is often boldly assumed
that
  $\displaystyle{ \lim_{T\to 0}S = 0}$
as soon as the third law has been described. Even Sommerfeld could not prevent
from doing so\cite{Sommerfeld}. One may ask the author who did so: Dose
  $\displaystyle{ \lim_{T\to 0}S = 0}$
contradict with the assumption such as that the Helmholtz free energy $F$ is
a homogeneous function of the first degree in extensive variables $V$ and $n$,
\etc? If no, is such an assumption independent of the third law?

In fact,
  $\displaystyle{ \lim_{T\to 0}S = 0}$
is a corollary of such an assumption together with the third law: According
to eq.~(\ref{Sdef}), $S_0(V,n) = \displaystyle{\lim_{T\to 0} S(T,V,n)}$ is a
homogeneous function of $V$ and $n$, say. At the same time, it is a constant
as asserted by the third law of thermodynamics. Hence, using Euler's theorem,
we obtain
\begin{displaymath}
  S_0 = \frac{\partial S_0}{\partial V}\, V
   + \frac{\partial S_0}{\partial n}\, n
  = 0.
\end{displaymath}
To my opinion, the assumption that a particular thermodynamic potential is a
homogeneous function in its corresponding extensive variables is and
\emph{must be} independent of the third law. One may name such an assumption as
the fourth law of thermodynamics, if he likes to. As a consequence, \emph{the
belief that each of thermodynamic potentials as well as the entropy can be
determined only up to a constant\footnote{
 Strictly speaking, the potentials $F$ and $G$ both have two constants before
they are fixed. For example, let $S$ and $S' = S + S_0$ describe the same
system in the same state, and let the internal energies $U$ and $U' = U + U_0$
describe also the same system in the same state. Then the corresponding
Helmholtz free energies will be $F = U - TS$ and
  $F' = U' - TS' = F + U_0 - T S_0$,
respectively. The same thing happens in the Gibbs free energy:
  $G' = G + U_0 - T S_0$.
Rigorously, all the potentials can be determined up to at most two constants
before they are fixed: One comes from the internal energy, and the other comes
from the entropy.
} is incorrect.}

\subsection{The Entropy and the Heat}
\label{sec:Heat}

How to calculate the heat that is absorbed by an open system? If the process is
reversible, then, according to thermodynamics, the heat absorbed by the open
system is $\dbar Q = T\,\dd S$ where $\dd S$ is the total differential of the
entropy of that open system. The answer is so simple that it seems absurd
to ask such a question. However, the following analysis indicates that the
so-called adiabatic process, which is a reversible process with the entropy $S$
remaining to be constant, is not a process as we have expected from the word
adiabatic.

First let us consider a simple example.

Suppose that there is a fluid system consists of one kind of molecules amounted
as $n_0$. The system is in equilibrium at a temperature $T$ and a volume $V_0$.
Then, according to the zeroth law of thermodynamics, there should be no heat
transferred from one part of it to another. Meanwhile, the internal energy of
this system is $U_0 = - pV_0 + TS_0 + \mu n_0$ where $S_0$ is the entropy of
the system, as indicated by eq.~(\ref{Uidentity}).

Now let us consider a subsystem of it which has a virtual boundary that
separates the subsystem from the whole. Suppose that the virtual boundary of
the subsystem expands slowly such that the center of mass remains unchanged and
the process can be considered as reversible. The subsystem is obviously an open
system, whose state is denoted by $(T,V,n)$ with an entropy $S=S(T,V,n)$.

Since the virtual boundary and the process are just imaginary, the intensive
quantities $p$, $T$ and $\mu$ are constant. Assume the initial and the final
states of the subsystem are
  $(T, V_1, n_1)$ and $(T, V_1 + \Delta V, n_1 + \Delta n)$,
respectively, then
\begin{equation}
  \frac{\Delta V}{V_0} = \frac{\Delta S}{S_0} = \frac{\Delta n}{n_0}.
\label{virtualDelta}
\end{equation}  
Hence, according to the formulism of thermodynamics, certain amount of heat
$Q=T\ \Delta S\neq 0$ must be absorbed by the subsystem after its virtual
expansion. In other words, as a virtual closed surface moves in an equilibrium
system, one has to admit that certain an amount of heat as $\dbar Q= T\,\dd S$
has been transferred through it, \emph{accompanying} the substance that goes
through the surface, even though the system is in equilibrium.

Then questions arise immediately. First, we have seen that the above subsystem,
separated by a virtual surface, does absorb some heat if its volume has been
changed. On the other hand, however, the zeroth law of thermodynamics should
have repelled such a possibility, because heat will not be transferred between
two systems or two parts of a system provided there is no difference of
temperature. Secondly, ever since the first law of thermodynamics was accepted,
heat has been treated as a form of energy transfer, being no longer anything
accompanying the ordinary substance as that was insisted in the caloric theory.
Now it seems that the caloric theory is somehow restored in thermodynamics.

So, there are serious questions in thermodynamics. We leave these questions to
be discussed in \cite{Zhou}. However, we must make our view points to these
questions clear before we perform further discussions. 

(1) For a system, the heat absorbed by it in a reversible process is always
calculated to be $\dbar Q = T\,\dd S$, no matter whether it is an open system
or not. To say a reversible process is adiabatic, it is equivalent to say that
$\dd S = 0$ in the process. As a consequence, in order that a reversible
process is adiabatic if certain a number of molecules have been brought into
the system, it must emit a suitable quantity of heat into the surroundings,
because the newly added molecules always tend to increase the total entropy of
the open system. Therefore the concept of adiabatic processes is far away from
what we have expected intuitively. \emph{As for how a process can be identified
to be adiabatic in the experiments, it is not the topic of this paper.}

(2) For simplicity, we only discuss open systems consist of only one kind of
molecules. When the amount $n$ of the molecules in the open system are changed,
the contribution of the newly added molecules to the entropy has been included
in $\dd S$, because $S$ depends on $n$, too, as shown in eq.~(\ref{Sdef}).
If the amount of molecules in the system remains constant, the change of
entropy in a reversible process is given by
\begin{displaymath}
  \bigg(\frac{\partial S}{\partial T}\bigg)_{V,n}\dd T
  + \bigg(\frac{\partial S}{\partial V}\bigg)_{T,n}\dd V.
\end{displaymath}
For a reversible process in which the amount $n$ is changing, the contribution
of the added molecules is
\begin{displaymath}
  \bigg(\frac{\partial S}{\partial n}\bigg)_{T,V}\dd n,
\end{displaymath}
and the total differential of the entropy, $\dd S$, is the sum of the above two
parts. In terms of $\dd S$, the contribution of the added molecules needs not
to be considered any more.

As an example, let us consider the fluid system with an open subsystem enclosed
by a virtual boundary, as introduced formerly. When the virtual boundary is
changed, the total change of the entropy of the subsystem consists of two parts:
\begin{displaymath}
  \bigg(\frac{\partial S}{\partial V}\bigg)_{T,n} \dd V
  + \bigg(\frac{\partial S}{\partial n}\bigg)_{T,V}\dd n.
\end{displaymath}
Due to the fact that $\frac{\dd V}{V} = \frac{\dd n}{n}$ where $V$ and $n$ are
the volume and the amount of molecules of the subsystem, respectively, one needs
not to know the details of the entropy function, yielding
\begin{eqnarray*}
  \bigg(\frac{\partial S}{\partial V}\bigg)_{T,n} \dd V
  + \bigg(\frac{\partial S}{\partial n}\bigg)_{T,V}\dd n
  & = & \bigg(\frac{\partial S}{\partial V}\bigg)_{T,n}\frac{V}{n}\,\dd n
  + \bigg(\frac{\partial S}{\partial n}\bigg)_{T,V}\dd n
\\
  & = & \bigg[\, V \bigg(\frac{\partial S}{\partial V}\bigg)_{T,n}
  + n  \bigg(\frac{\partial S}{\partial n}\bigg)_{T,V}\, \bigg]\,
  \frac{\dd n}{n}
\\
  & = & \frac{S}{n}\,\dd n = \frac{S}{V}\,\dd V.
\end{eqnarray*}
Since both $\frac{S}{n} = \frac{S_0}{n_0}$ and $\frac{S}{V} = \frac{S_0}{V_0}$
remain constant in the process, we can integrate the above to give
\begin{displaymath}
  \Delta S = \frac{S_0}{n_0}\,\Delta n = \frac{S_0}{V_0}\,\Delta V,
\end{displaymath}
coinciding with eqs.~(\ref{virtualDelta}). Obviously, if one is about to
consider the contribution of the molecules that flows into the subsystem once
again, he is definitely doing something wrong.

(3) Is heat a kind of substance accompanying ordinary matter, as it was said
in the old-fashioned caloric theory, or just the energy transferred from one
body to another as the result of the difference of temperature? As we know,
thermodynamics has chosen the latter. But the following consideration will
make thermodynamics to sink into trouble.

As what happens in $\dd S$, whenever a process is reversible, the heat
\begin{eqnarray}
  \dbar Q & = & T\dd S
  = T\bigg(\frac{\partial S}{\partial T}\bigg)_{V,n} \dd T
  + T \bigg(\frac{\partial S}{\partial V}\bigg)_{T,n}\dd V
  + T \bigg(\frac{\partial S}{\partial n}\bigg)_{T,V}\dd n
\nonumber \\
  & = & C_{V,n}\,\dd T + T \bigg(\frac{\partial S}{\partial V}\bigg)_{T,n}\dd V
  + T \bigg(\frac{\partial S}{\partial n}\bigg)_{T,V}\dd n
\end{eqnarray}
consists of two parts: the ordinary part
  $C_{V,n}\,\dd T + T \big(\frac{\partial S}{\partial V}\big)_{T,n}\dd V$
and the heat ``brought into the system by $\dd n$",
  $T \big(\frac{\partial S}{\partial n}\big)_{T,V}\dd n$.
The latter makes the heat something like the companion of the ordinary
substances. Thus, reasonably, some of us will feel uncomfortable because the
spirit of the caloric theory is found wandering in thermodynamics. But we do
not want to discuss it, but leaving it into \cite{Zhou}.

\section{How to Measure the Entropy}
\label{sec:Method}

\subsection{The Outline of the Measuring Methods}
\label{sec:Outline}

For simplicity we only consider the gas systems, each of which consists of
only one kind of molecules. For such a system, its equilibrium state can be
specified by three variables, the absolute temperature $T$, the volume $V$,
and the amount $n$ of the molecules, say. Then there are the equations of
state for such a system. One of these equations is
  $ p = p(T,V,n), $
which can be obtained state by state. The partial derivatives of $p$ with
respect to $T$, $V$ and $n$, respectively, can also be measured state by state.

In order to measure the entropy $ S = S(T,V,n) $ of such a system in a given
state $(T,V,n)$, suppose that there is a reversible
isothermal process
\begin{equation}
  T = \mathrm{constant}, \qquad V = V(n).
\label{process}
\end{equation}
Let $(T, V, n)$ and $(T,V + \dd V, n + \dd n)$ be the states in the above
process, with $\dbar Q$ the heat that is absorbed by the system. Then the
entropy of the system in the state $(T,V,n)$ is
\begin{equation}
  S(T,V,n) = \frac{n}{T}\,\frac{\dbar Q}{\dd n}
  + \bigg(\frac{\partial p}{\partial T}\bigg)_{V,n}
    \bigg( V - n \frac{\dd V}{\dd n} \bigg).
\label{method1}
\end{equation}
Therefore the entropy can be obtained provided that $\dbar Q$ has been
obtained. Especially, if the process is both isothermal and adiabatic, namely,
$T = \mathrm{constant}$ and $\dbar Q = 0$, the entropy in the state $(T,V,n)$
will be
\begin{equation}
  S(T,V,n) = \bigg(\frac{\partial p}{\partial T}\bigg)_{V,n}
  \bigg( V - n\,\frac{\dd V}{\dd n}\bigg).
\label{isoad}
\end{equation}

\subsection{The Principles of the Experiment}
\label{sec:ProposalExp}

As we know, for a gas system, the first law of thermodynamics can be written in
the form of
\begin{displaymath}
  \dd U = - p \,\dd V + T \,\dd S + \mu \,\dd n.
\end{displaymath}
A Legendre transformation gives the differential of the free energy as
\begin{equation}
  \dd F = - p \,\dd V - S \,\dd T + \mu \,\dd n,
\label{dF}
\end{equation}
where the free energy is defined as
$
  F = U - TS.
$
Being a homogeneous function of the first degree in $V$ and $n$ for any given
temperature $T$, Euler's theorem for homogeneous functions of the first
degree must be satisfied, in which the temperature is treated as a parameter,
giving
\begin{equation}
  F = - pV + \mu n.
\label{F}
\end{equation}
The optional form of the above equation can be
$
  U = - pV + TS + \mu n
$
for the internal energy, or\cite{Zemansky}
$
  G = \mu n
$
for the Gibbs free energy, as what have been mentioned in
\S\ref{sec:AbsEntropy}.

One of the most important consequences of the above equations is that the
intensive quantities $p$, $T$ and $\mu$ are not functionally independent,
namely, their differentials are constrained by an equation
\begin{equation}
  \dd \mu = \frac{V}{n}\,\dd p - \frac{S}{n}\,\dd T.
\label{diffmu}
\end{equation}
In fact, it can be obtained by the comparison of the differential of
eq.~(\ref{F}) with eq.~(\ref{dF}). The above equation implies that the chemical
potential $\mu$ is a function of the intensive variables $p$ and $T$. Since the
pressure $p$ can be viewed as a function of the variables $T$, $V$ and $n$, we
can obtain
\begin{equation}
  \bigg(\frac{\partial\mu}{\partial T}\bigg)_{V,n}
  = \frac{V}{n}\bigg(\frac{\partial p}{\partial T}\bigg)_{V,n}
  - \frac{S}{n}
\end{equation}
by using the chain rule of partial derivatives.

By virtue of the above equation, the heat that is absorbed by an open system in
a reversible process reads
\begin{equation}
  \dbar Q = T \,\dd S = C_{V,n}\,\dd T
    + T \bigg(\frac{\partial p}{\partial T}\bigg)_{V,n}\dd V
    + \frac{T}{n}\,\bigg[ S - V \bigg(\frac{\partial p}{\partial T}\bigg)_{V,n}
      \bigg]\,\dd n,
\label{dbarQ}
\end{equation} 
in which the partial derivatives of the entropy $S$ with respect to $V$ and $n$,
respectively, have been replaced by the partial derivatives of $p$ and $-\mu$,
both with respective to $T$, according to Maxwell's relations
\begin{equation}
  \bigg(\frac{\partial S}{\partial V}\bigg)_{T,n}
  = \bigg(\frac{\partial p}{\partial T}\bigg)_{V,n},
\qquad
  \bigg(\frac{\partial S}{\partial n}\bigg)_{T,V}
  = - \bigg(\frac{\partial \mu}{\partial T}\bigg)_{V,n}.
\label{MaxwellEqs}
\end{equation}
As we know, the heat capacity $C_{V,n}$ in eq.~(\ref{dbarQ}) is defined as
\begin{displaymath}
  C_{V,n} = T\bigg(\frac{\partial S}{\partial T}\bigg)_{V,n}.
\end{displaymath}

For an isothermal reversible process (\ref{process}), we obtain the equation
\begin{displaymath}
  \frac{\dbar Q}{\dd n}
  = T\,\bigg(\frac{\partial p}{\partial T}\bigg)_{V,n}\, \frac{\dd V}{\dd n}
  + \frac{T}{n}\,\bigg[\, S - V\,\bigg(\frac{\partial p}{\partial T}\bigg)_{V,n}
    \bigg]
\end{displaymath}
from eq.~(\ref{dbarQ}). Hence eq.~(\ref{method1}) can be obtained. If the
process (\ref{process}) is not only isothermal, but also adiabatic, then
$\dbar Q = 0$. Thus eq.~(\ref{isoad}) can be obtained.

\section{Conclusions and Discussions}
\label{sec:cd}

In \S\ref{sec:AbsEntropy}, it was concluded that the concept of abstract
entropy makes sense in thermodynamics. So there should be some methods to
measure the entropy. The method presented in this paper is one of them.

As we have indicated in \S\ref{sec:Method}, were it possible that
(1) we can design an infinitesimal reversible process that is isothermal with
the amount $n$ of molecules being changed and that
(2) we can measure the heat absorbed by the open system in this process, then
the entropy of the system can be measured. If,
(3) in addition, the infinitesimal process is adiabatic, it could not be better.

However, we have seen in \S\ref{sec:Heat} that the concept of heat in
thermodynamics is not so simple as it looks. It seems to be a question whether
we can measure the heat if, in the process, any one of the amounts of the
components is changed. This is due to the fact, in thermodynamics, that the
molecules that are brought into or out of an open system are accompanied with
certain a quantity of heat, as shown in the example of the subsystem of an
equilibrium fluid system (see, \S\ref{sec:Heat}). As a consequence, it can be
asked whether the method presented in this paper is practical.

We must notice that, in order to measure the entropy $S(T,V,n)$ in a given
state $(T,V,n)$, one process satisfying the conditions (1) and (2) in the
above is sufficient. If there is no such a reversible process, we may refer
to thermodynamics as a theory speaking of nothing. As long as we believe in
thermodynamics, we have to admit that the method of measuring the entropy is
reasonable and practical.

It is in this sense that the entropy can be measured.

\vskip 2cm
\begin{center} \textbf{\large Acknowledgments} \end{center}

The author wants to thank Prof. H. Y. Guo, Prof. Z. Zhao and Prof. H. Y. Wang
for helpful discussions on such topics. He is also thankful to Dr. Xin Wang,
Dr. Yu-Guang Wang, and Dr. Lian-You Shan for stimulating discussions. Special
appreciations are given to Ji-Jun Li, whose effort on thermodynamics several
years ago is a historical background as why the author paid his attention to
these topics.

\end{document}